%% file: paper.tex
\RequirePackage{snapshot}
\documentclass{llncs}

\usepackage{amsmath}
\usepackage{amssymb}
\usepackage{graphicx}
\usepackage{subfig}

\usepackage{color}

\usepackage{algorithm}
\usepackage{algpseudocode}

\title{Coloured and task-based stencil codes}

\titlerunning{Colouring and task-based stencil codes}

\author{
  Benjamin Hazelwood
  \and
  Tobias Weinzierl
  \thanks{
    This work received funding from the European Union’s Horizon 2020
    research and innovation programme under grant agreement No 671698
    (ExaHyPE).
    It made use of the facilities of the Hamilton HPC Service of Durham
    University.
    We appreciate the support of the RSC Group by granting us access to their
    KNL nodes.
  }
}

\institute{
  School of Engineering and Computing Sciences\\
  Durham University\\
  Great Britain \\
  \email{{\{benjamin.hazelwood,tobias.weinzierl\}}@durham.ac.uk}
}

\begin{document}
\maketitle

\begin{abstract}
  \input{00_abstract}

\end{abstract}

\input{01_introduction}

\input{02_stencil}
\input{03_algorithms}
\input{05_results}
\input{06_insights}

\input{09_conclusion}

\bibliographystyle{splncs03}
\bibliography{paper}


\end{document}

%% file: 00_abstract.tex
Simple stencil codes are and remain an important building block in scientific
computing.
On shared memory nodes, they are traditionally parallelised through 
colouring or (recursive) tiling. 
New OpenMP versions alternatively allow users to specify data
dependencies explicitly and to outsource the decision how to distribute the work
to the runtime system.
We evaluate traditional multithreading strategies on both Broadwell and KNL,
study the arising assignment of tasks to threads and, from there,
derive two efficient ways to parallelise stencil codes on regular Cartesian
grids that fuse colouring and task-based approaches.

%% file: 01_introduction.tex
\section{Introduction}

%
%
Stencil codes applying a fixed data access pattern over a
mesh are an important building block of many simulation codes.
We find them, for example, within smoothers of matrix-free multigrid
codes or as fundamental building block of explicit time stepping schemes.
While they can be generalised to arbitrary meshes, we use the term stencil
\cite{Demmel}
in the context of regular (topologically) Cartesian grids as they
arise for example in block-structured \cite{Dubey:14:AMRFramework} codes
(Fig.~\ref{figure:stencils:amr}).
Since they are often algorithmic workhorses
which make up for a significant share of the total runtime, it is
important to make the stencil evaluations fast and scale.

%
%
Though mainstream compilers and programming languages lack built-in support for
stencil expressions, modern chips grow into high-throughput, streaming architectures
that are by construction well-suited for stencil evaluations:
vector registers widen, caches appreciate regular data accesses, arithmetic
units may combine multiplications and additions.
One might read these trends as a stencil-aware/-friendly hardware evolution
\cite{Dongarra:14:ApplMathExascaleComputing}.
However, as architectures become more powerful, they also tend to become
more sensitive to synchronisation as it is induced by bulk synchronous
programming (BSP) \cite{BSP} or work stealing.
From time to time, we thus have to re-evaluate our patterns how to program
stencil codes efficiency.
On the software side, stencil compilers
\cite{PlutoScheduler,Exastencils}
step in to mitigate and exploit the growing hardware
complexity.
They typically augment stencil expressions by data layout
and traversal instructions which offers more opportunities to rearrange evaluations
aggressively.
They widen the optimisation space.
At the same time, libraries, programming interfaces and programming styles continue to improve.
Notably, task-based parallelism becomes popular.
Here, stencil dependencies are explicitly specified, and the stencil
evaluation ordering is outsourced to a runtime system rather than falling into
the responsibility of a (stencil) compiler.

%
%
While dedicated stencil compilers are one path towards
upscaling, simple stencil loops with simple traversal orders not
exploiting or benefiting from domain-specific stencil compilers will remain
omnipresent in research codes.
Also, techniques such as higher-order Discontinuous Galerkin methods which
yield stencil-like access patterns yet are difficult to rephrase in stencil notation will gain
importance \cite{Dongarra:14:ApplMathExascaleComputing}.
Programmers therefore will continue to optimise stencil expressions manually
and, with the advent of task systems and manycore architectures,
question common wisdom how to manually write fast stencil codes.

%
%
We focus on four particular questions here.
First, is it true that colouring approaches \cite{Yeo} that resolve
stencil data races a priori are the best choice for small stencils and meshes,
or are domain decomposition approaches such as nested dissection
\cite{George73,Kumar12} superior as they exhibit fewer global synchronisation
points? Second, does the advent of OpenMP's task dependency mechanism
\cite{OpenMP4:2013} render classic stencil techniques
obsolete? Third, do flexible task-based formalisms really yield new execution patterns or
do they automatically fall back to execution orders that we know from
traditional manual optimisation?
Finally, what are efficient and elegant programming techniques yielding
performance through new language features?

%
%
Our studies are organised as follows:
After a brief introduction of the stencils used,
we introduce our testbed of algorithmic building blocks in Section \ref{section:algorithms},
before we revise the studied, traditional parallelisation approaches.
The main part of the present paper discusses the observed execution
characteristics (Section \ref{section:results}) from which we derive for our
setups how new OpenMP might influence efficient programming
(Section \ref{section:insights}).
A brief summary and an outlook close the discussion.

%% file: 02_stencil.tex
\section{The studied stencil codes}
\label{section:stencils}

\begin{figure}[htb]
 \begin{center}
  \begin{minipage}[l]{0.4\textwidth}
    \includegraphics[width=0.95\textwidth]{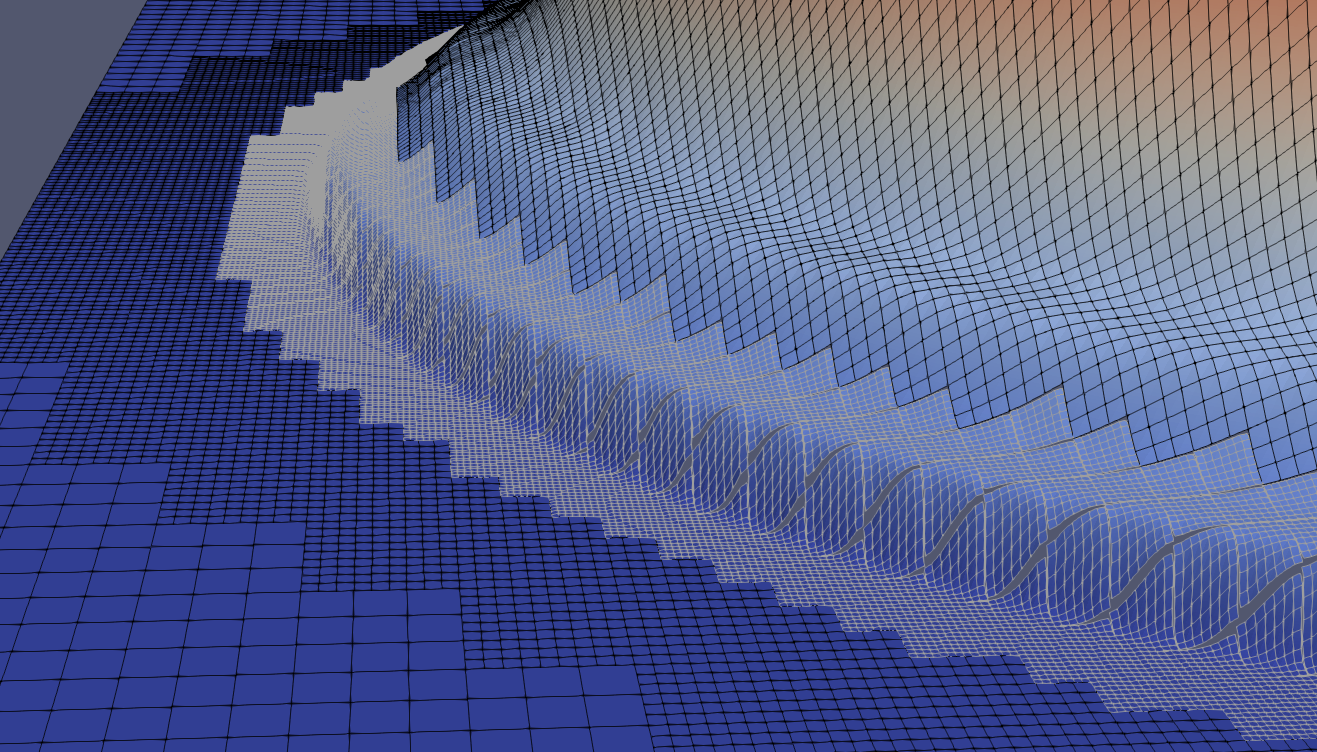}
  \end{minipage}
  \begin{minipage}[r]{0.4\textwidth}
   \[
    \left[
     \begin{array}{rrr}
      0 & -1 & 0 \\
      -1 & 4 & -1 \\
      0 & -1 & 0
     \end{array}
    \right]
    \qquad
    \left[
     \begin{array}{rrr}
      -\frac{1}{3} & -\frac{1}{3} & -\frac{1}{3} \\
      -\frac{1}{3} & \frac{8}{3} & -\frac{1}{3} \\
      -\frac{1}{3} & -\frac{1}{3} & -\frac{1}{3}
     \end{array}
    \right]
   \]
  \end{minipage}
 \end{center}
 \vspace{-0.4cm}
 \caption{
  Left: example for patch-based adaptive grid of a Finite Volumes solver from
  the ExaHyPE project.
  Middle and right: The classic 5-point Finite Differences and 9-point Finite
  Element stencil in two dimensions.
 }
 \label{figure:stencils:amr}
\end{figure}

%
%
The present manuscript focuses on two- and three-dimensional Cartesian meshes.
They are modelled as plain $d$-dimensional arrays.
If not used directly as solver data structure, we find such data structures as
patches in patch-based adaptive mesh refinement \cite{Dubey:14:AMRFramework}.
As our studies rely on a topological regularity
(Fig.~\ref{figure:stencils:amr}), they can be generalised to unstructured meshes hosting regular patches.
The analysis of dynamically changing, adaptive meshes however is beyond scope.

%
%
The algorithmic blueprint of our objects of studies is simple:
Given is a $d$-dimensional Cartesian mesh of fixed size.
Each mesh cell holds one scalar value.
We run over all cells of this mesh.
Per cell, we compute one value according to a fixed formula accepting the
cell's value and the neighbouring values, i.e.~according to a stencil, and we
write this value in-situ into the respective mesh cell.
This mirrors a Gau\ss-Seidel update of unknowns.
A separate output array, as used to determine a matrix-vector product, would
change all performance data quantitatively but not qualitatively as the 
arithmetic intensity is recalibrated.
Our studies are interested in the time per cell update.
As we write back in-situ, i.e.~work solely with one big double array,
vectorisation for standard lexicographic enumeration is manually enforced,
i.e.~we switch from a Gau\ss-Seidel update scheme into a block-Jacobi
Gau\ss-Seidel scheme.

%
%
The present manuscript focuses on compact $2d+1$-point and $3^d$-point
stencils (Fig.~\ref{figure:stencils:amr}).
The computation per mesh cell accesses the face-connected neighbour cells or all
adjacent cells, respectively, and computes a weighted sum of these
contributions.
Such stencils are found in classic low-order Finite Elements and Finite
Difference schemes \cite{Abramowitz}.
We generalise the stencil evaluations as follows:
Each stencil evaluation $S(k)$ accepts a parameter $k$.
For $k=0$, we assume the stencil entries to be given.
For $k>0$, we make the stencil entry computation itself require $k$ floating
point operations.
%
%
Our motivation is that we want to mitigate the arithmetic intensity and
non-prescribed data access patterns of high order schemes that on-the-fly
compute the stencil entries from sophisticated formulae.
An example for such an approach are non-trivial Finite Volume codes where the
flux function's branching is determined by the solution or the ADER-DG
scheme \cite{Dumbser:14::ADERDG} including a limiter.
The stencil cost here depend on the presence of shocks which in turn
trigger a Finite Volume-based limiter.

%% file: 03_algorithms.tex
\section{Evaluated traversal orders}
\label{section:algorithms}

\begin{figure}[htb]
 \begin{center}
  \includegraphics[width=0.22\textwidth]{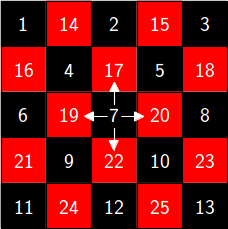}
  \hspace{0.8cm}
  \includegraphics[width=0.22\textwidth]{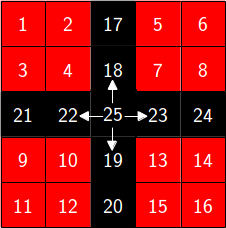}
 \end{center}
 \vspace{-0.4cm}
 \caption{
  Red-black colouring (left) of a two-dimensional $N \times N$ mesh with $N=5$
  and one-level nested dissection of the same mesh (right).
 }
 \label{figure:algorithms:colouring}
\end{figure}

We study three different orders how to run through all cells.
The first two approaches (Fig.~\ref{figure:algorithms:colouring}) are classics
from linear algebra \cite{Demmel}, whereas the third approach does not prescribe
the order but defines constraints on the traversal order and delegates the
actual run-through to OpenMP's runtime system.

%
%
\paragraph{Colouring.} Our first approach realises classic red-black colouring
for the 5-point ($d=2$) and 7-point ($d=3$) stencil and $2^d$ colouring
otherwise \cite{Demmel,Yeo}.
For $c$ colours, we end up with $c$ passes over the grid though each pass
updates only around $N^d/c$ cells for a $N \times N$ ($d=2$) or $N \times N
\times N$ ($d=3$) patch.
Each pass fixes one cell that has not been updated yet and from hereon skips
iteratively all those cells within the grid whose values depend on updated
cells.
The cell updates within each pass are embarrassingly parallel.
Typically, we have one loop over $c$ hosting, depending on the dimension, two or
three loops where the outer of these loops is parallelised with a
\texttt{parallel for}.
We may assume that the \texttt{parallel for} splits up the cells equally.
With a sequence of $c$ \texttt{parallel for}s, we end up with a classic, flat
(not nested) BSP-style algorithm.

%
%
\paragraph{Nested dissection.} Our second approach is nested dissection
\cite{George73,Kumar12} which is straightforward here as we restrict to compact
stencils and regular meshes \cite{Khaira92}.
We split the mesh into $2^d$ (roughly) equally-sized blocks separated by a thin
layer of one cell width and process the blocks concurrently.
There are no dependencies between them.
Once this is done, the algorithm handles the thin separation layer.
An recursive extension of the scheme is straightforward.
While one could recurse until any subregion consists of $2^d$ or fewer cells, we
recurse until the number of small concurrent blocks exceeds or equals the number
of threads, and then recurse once more to render work stealing possible.

The traversal's initial concurrency
exceeds the number of threads, but it reduces afterwards.
Each synchronisation point in the algorithm affects $2^d$ threads.
The
treatment of the coarsest separation layer implicitly synchronises globally.
There is no global synchronisation otherwise.
Nested dissection generalises BSP through a fork-join model.
It uses nested, hierarchical BSP.
In OpenMP, we can directly realise the algorithm through the \texttt{task}
construct.
While the total number of synchronisation points is in general higher per step
than for colouring, each synchronisation point affects only a small, fixed
number of threads.

%
%
\paragraph{Task-based modelling with dependency graphs.} Our last approach
starts from a task-based point of view but does not impose a task
execution ordering explicitly.
Instead, we rely on OpenMP's
superscalar tasks \cite{Agullo,Badia,Dagum}:
The evaluation of one cell's operator is a task that can, in principle, be ran
in parallel to any other cell task.
However, no two cells neighbouring each other may be processed at the same time.
A symmetric graph anticipating the stencil structure and thus being a regular
Cartesian graph itself formalises this.
With recent OpenMP versions, we can directly describe these \texttt{task}s
and model our constraints as \texttt{in} and \texttt{out} dependencies.
The memory location of a cell hereby uniquely defines the tasks.

A task-based formalism exposes the maximum concurrency possible to the
threading subsystem.
With work stealing in place, we may thus assume that the work is perfectly
balanced.
There is no explicit overall synchronisation at all, but a semaphore mechanism
in OpenMP has to ensure that no two tasks accessing neighbouring cells are ran
concurrently.
Nested dissection and colouring are one of many instantiations of a traversal
order that fits to the dependency graph.
Delegating the identification of a traversal order to a scheduler, we
however rely on this scheduler to puzzle out an efficient ordering on-the-fly.

%% file: 05_results.tex
\section{Measurements}
\label{section:results}

Our experiments were conducted on a cluster with Intel E5-2650V4 (Broadwell)
nodes with 24 cores per node that run at 2.20 GHz.
We have
256 KB L2 cache
per core and 30 MB smart cache shared between all cores.
AVX 2.0 is available.
Furthermore, we reran the experiments on an Intel Knights Landing
chip (Xeon Phi 7250) at 1.40 GHz.
A node hosts 68 cores which
share their 34 MB L2 cache.
AVX-512 is available.

The codes have been translated with the Intel compiler 17.0.2, and we have
validated through the vectorisation reports that SIMD facilities are exploited.
For the standard, cheap stencils ($k=0$) we hard-coded all stencil entries.
For the homogeneous, expensive stencil evaluations, we mitigate computations
by adding $k=100$ evaluations of the \texttt{sin} function to the prescribed
stencil entry.
For heterogeneous load patterns we make $k$ run linearly from one to 100
over the mesh, i.e.~1/100 of the cells use the standard stencil, 1/100 add one
\texttt{sin} evaluation to the entries, and so forth.
Though we focus on single grid sweeps, we do run a few thousand
sweeps to eliminate measurement inaccuracies.
Shared memory parallelisation relies on Intel's OpenMP.
We configure it through \texttt{OMP\_NUM\_THREADS} and delegate all pinning to
the scheduler (Slurm) with \texttt{OMP\_PROC\_BIND} switched on, i.e.~the
scheduler may not move threads between cores.

\begin{figure}[htb]
 \begin{center}
  \includegraphics[width=0.4\textwidth]{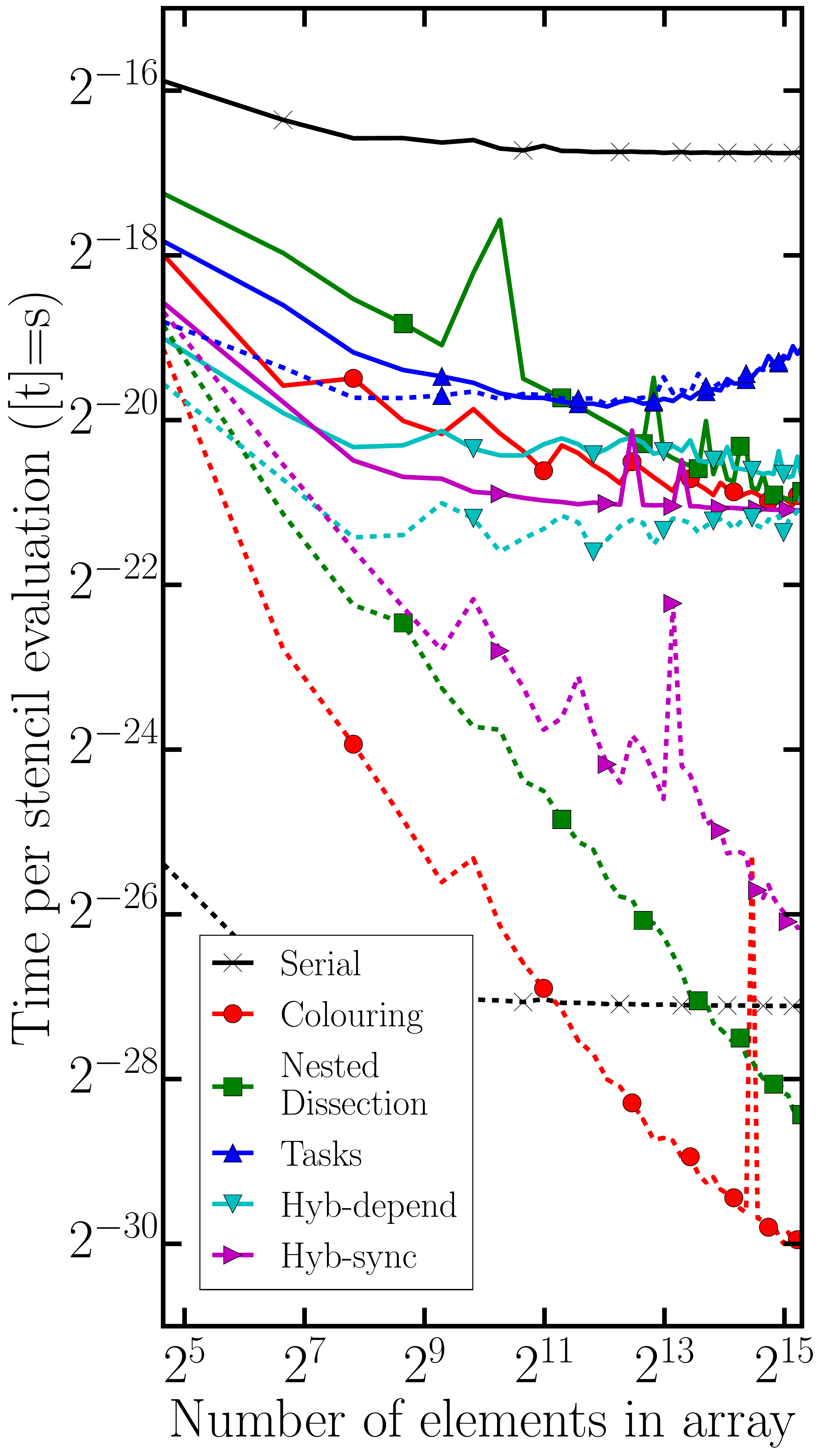}
  \hspace{0.6cm}
  \includegraphics[width=0.4\textwidth]{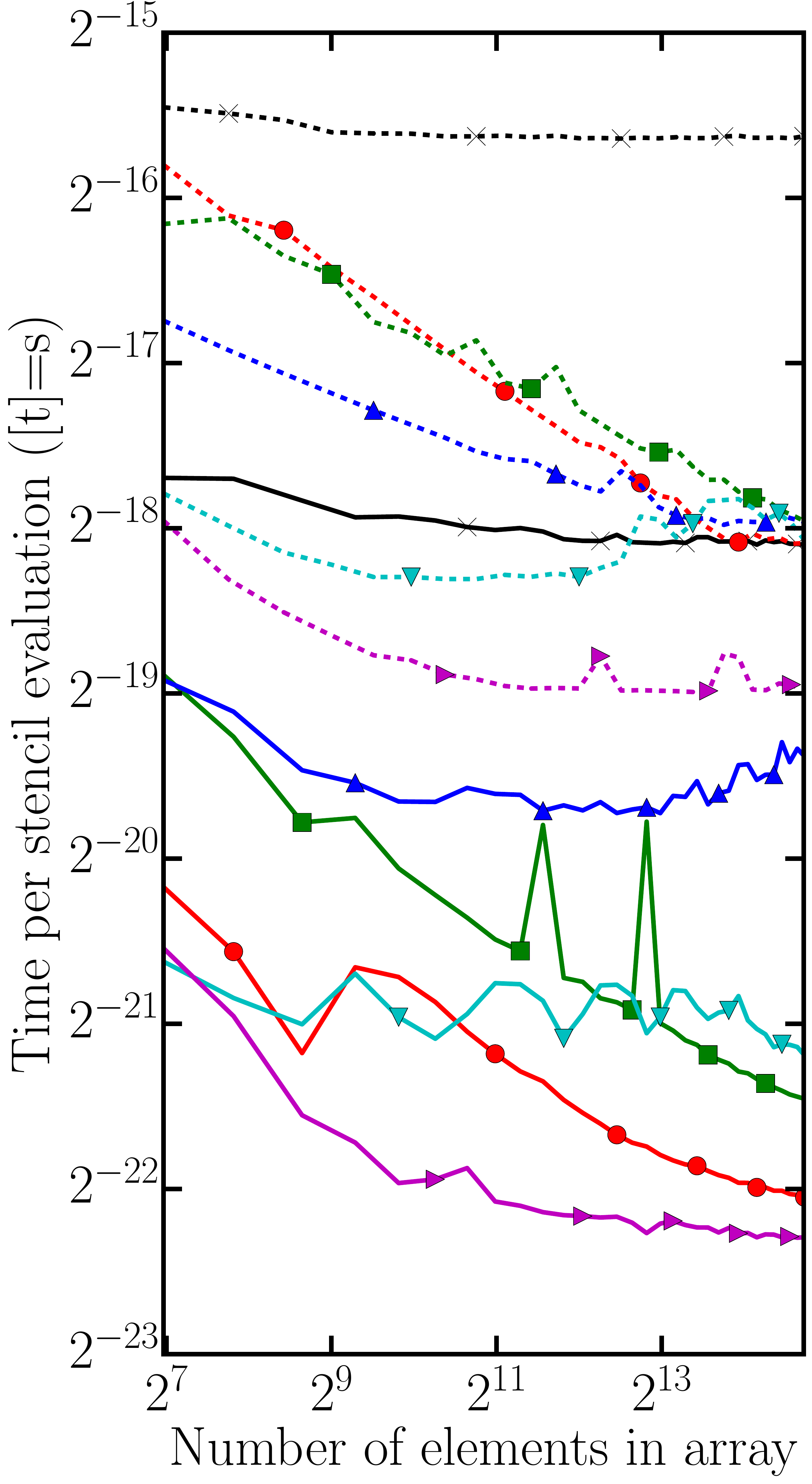}
 \end{center}
 \vspace{-0.6cm}
 \caption{
   Broadwell experiments. Left: We run a fixed
   5-point stencil (dotted) or invoke around 100 \texttt{sin} operations per run
   (solid) on a two-dimensional grids.
   Right: Non-uniform number of \texttt{sin} evaluations
   for the 5-point stencil (solid) or the three-dimensional 27-point stencil
   (dotted).
 }
 \label{figure:results:broadwell}
\end{figure}

\begin{figure}[htb]
 \begin{center}
  \includegraphics[width=0.4\textwidth]{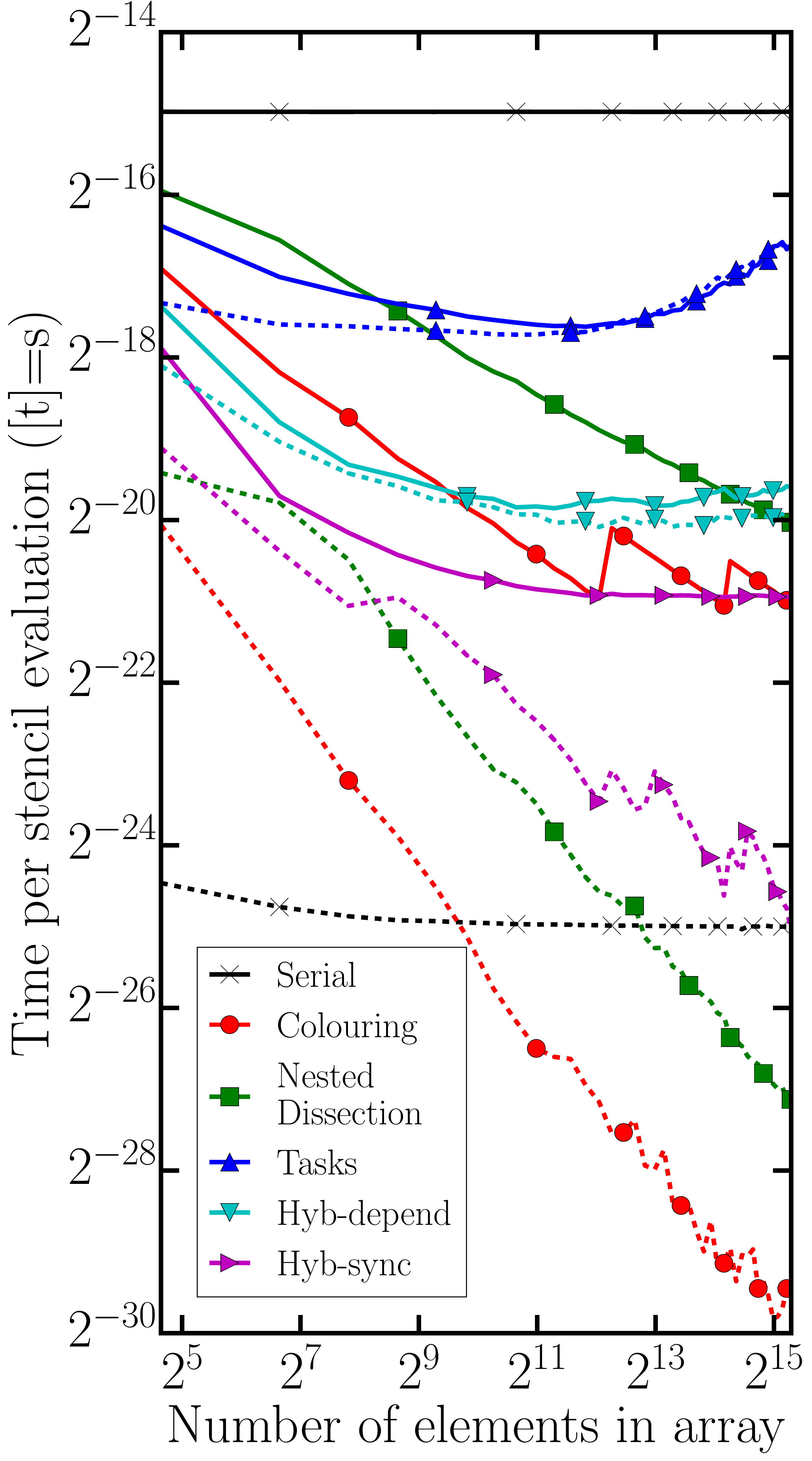}
  \hspace{0.6cm}
  \includegraphics[width=0.4\textwidth]{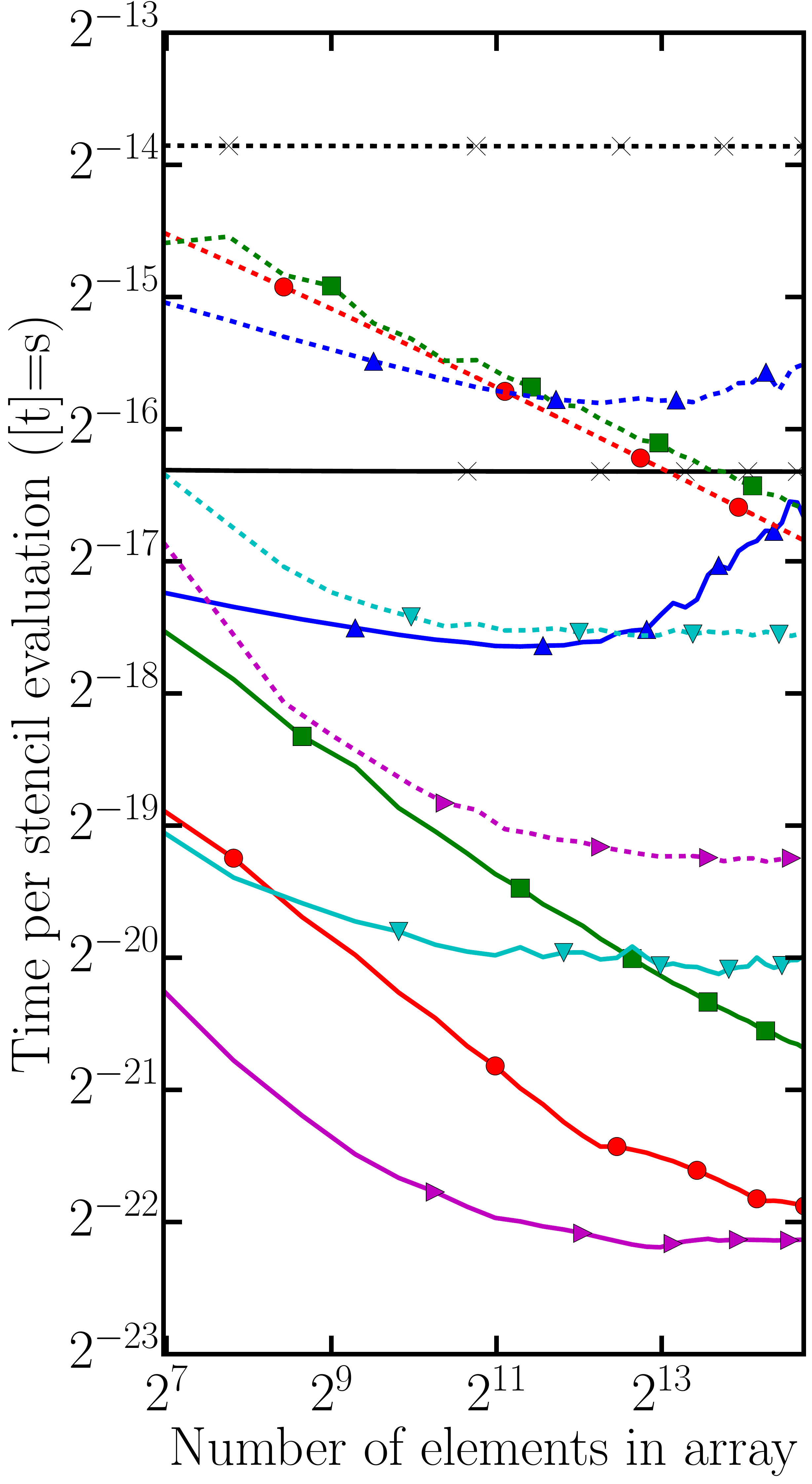}
 \end{center}
 \vspace{-0.6cm}
 \caption{
   Experiments from Fig.~\ref{figure:results:broadwell} reran on the KNL.
 }
 \label{figure:results:knl}
\end{figure}

%
%
We observe that all schemes suffer from overhead.
As a result, the multithreaded code variants perform worse than a pure
serial run if the arithmetic load is small, homogeneous and the problem is tiny
(Fig.~\ref{figure:results:broadwell} and \ref{figure:results:knl}).
On the Broadwell, a tie is observed for $2^{11}$ array elements (cells) the
earliest, on the KNL we run into the tie at $2^{10}$.
Grids have to exceed $10\times 10$ to benefit from threading.
Colouring yields by far the best execution times.
Nested dissection comes second but the gap between the two of them widens.
Tasking provides some speedup for tiny grids but then soon the speedup flattens
or even deteriorates.

%
%
Colouring continues to be faster than nested dissection for expensive
homogeneous stencil evaluations.
For both, multithreading overhead is amortised immediately.
Nested dissection starts to yield the lowest performance but, with increasing
mesh sizes, closes the gap to colouring and eventually runs into a tie.
The dependency task concept yields speedup, but again deterioriates if we make
the mesh bigger.
In two dimension, is starts off in-between nested dissection and
colouring and thus is inferior to colouring.
In three dimensions, it starts off better than colouring and remains faster than
colouring for up to $2^{12}$ cells.

%
%
In two dimensions, the speedup observations continue to hold if the load per
stencil becomes inhomogeneous.
Enabling and using work stealing in all approaches---our colouring relies on
\texttt{OMP\_SCHEDULE} \texttt{dynamic}---does not make a
difference.
Things change in three dimensions where the number of dependencies per task
increases.
Nested dissection and colouring yield comparable throughputs, while tasking
outperforms the two of them as long as the problem size is reasonably modest.
As tasking still deteriorates, there is a mesh size where the other two
approaches catch up and overtake for sparse stencils with seven entries.
For the 27-point stencil, we end up with the same performance for all setups for
large grids.

 \begin{figure}[htb]
  \begin{center}
   \subfloat[Colouring]{\includegraphics[width=0.25\textwidth]{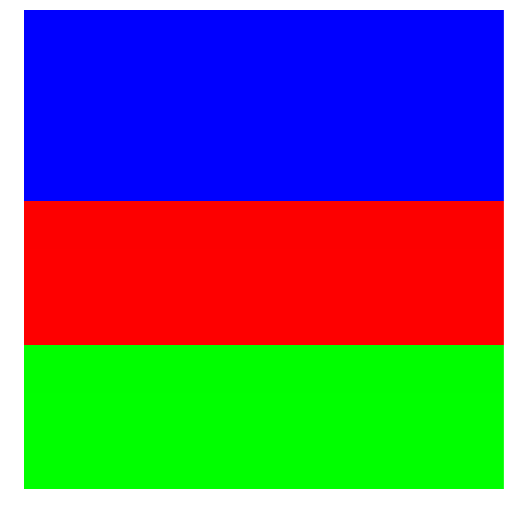}\label{fig:col-homo}}
   \qquad
   \subfloat[Nested dissection]{\includegraphics[width=0.25\textwidth]{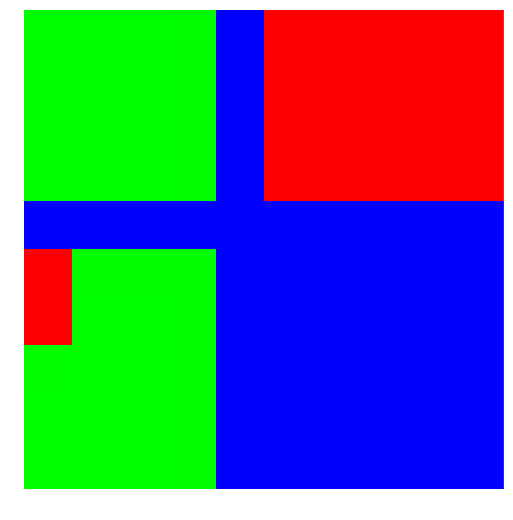}\label{fig:nd-homo}}
   \qquad
   \subfloat[Graph-based]{\includegraphics[width=0.25\textwidth]{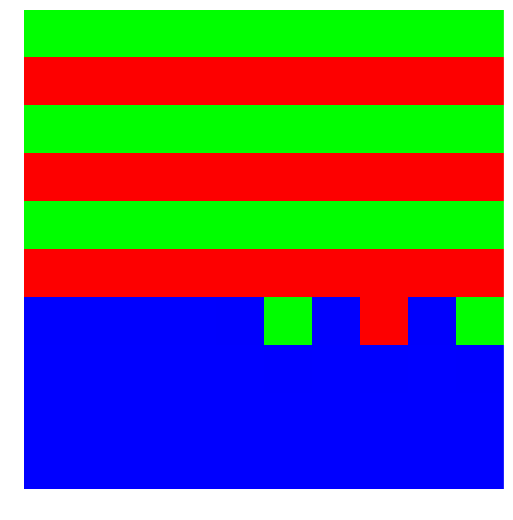}\label{fig:dyn-homo}}
  \end{center}
  \vspace{-0.6cm}
  \caption{
    Access pattern for the classic two-dimensional 5-point stencil ran on a
    $10\times 10 $ grid. Three threads are used and the colours illustrate which
    thread accesses which cell.
  }
  \label{figure:results:task-assignment-homogeneous}
 \end{figure}

 \begin{figure}[htb]
  \begin{center}
   \subfloat[Colouring]{\includegraphics[width=0.25\textwidth]{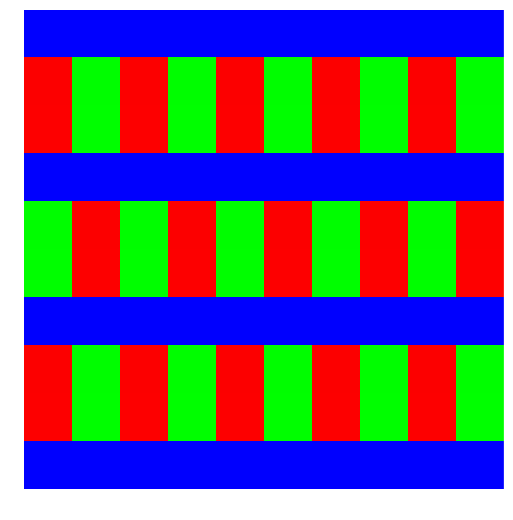}\label{fig:col-het}}
   \qquad
   \subfloat[Nested dissection]{\includegraphics[width=0.25\textwidth]{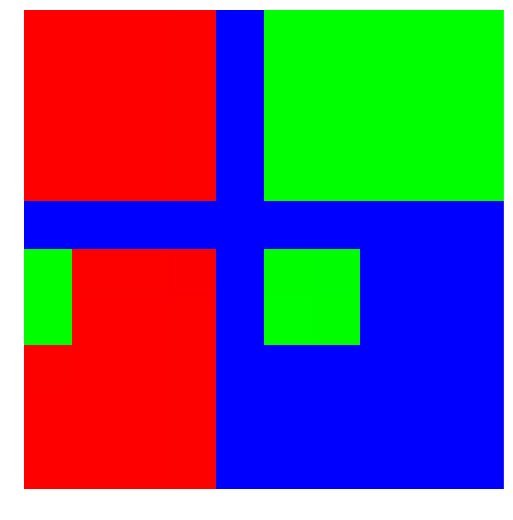}\label{fig:nd-het}}
   \qquad
   \subfloat[Graph-based]{\includegraphics[width=0.25\textwidth]{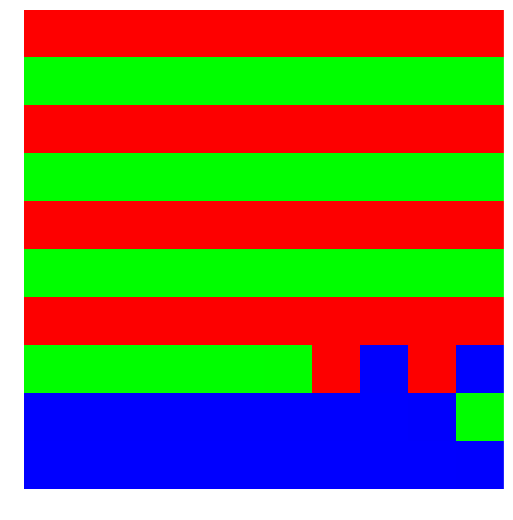}\label{fig:dyn-het}}
  \end{center}
  \vspace{-0.6cm}
  \caption{
   Experiments from Fig.~\ref{figure:results:task-assignment-homogeneous} reran
   for heterogeneous stencil loads.
   }
  \label{figure:results:task-assignment-heterogenous}
 \end{figure}

%
%
It is enlightning to study how the three concepts distribute the work.
Each colour tends the split up the iteration range roughly the same if the
computational load is homogeneous
(Fig.~\ref{figure:results:task-assignment-homogeneous}).
We implicitly obtain an affinity-aware data decomposition.
For heterogeneous stencils, the compiler collapses the spatial loops---our
coding runs top-down first---and splits them up in a colouring-alike pattern
(Fig.~\ref{figure:results:task-assignment-heterogenous}).
Only the first thread continues to hold lines of responsibility.

Nested dissection yields the expected data assignment patterns which clearly
illustrate its ill-balancing.
The relative difference of its responsibility decompositions to total workload
is the more significant the smaller the total workload.
It thus is not a surprise that the gap between colouring and nested dissection
closes the more the bigger the solved problem.
Nested dissection implicitly yields loop tiling.
We do not reuse data---we are interested in single sweep performance---and thus
cannot exploit the caches through tiling.
This would change if we studied multi-pass performance.

OpenMP's task dependency mechanism mixes the two assignment patterns:
The root thread spawns all other threads alternatingly along the outer loop.
From hereon the threads process their data line-wisely.
After a reasonable amount of data is initially deployed, the root thread itself
retains a whole block of work and starts to process it.
As the non-root threads start to work ahead, they later steal entries from this
master block.

%% file: 06_insights.tex
\section{Two alternative realisation variants}
\label{section:insights}

\algrenewcommand\algorithmicindent{0.6em}

\begin{algorithm}[htb]
  \caption{
    Algorithms \texttt{Hyb-depend} (left) and \texttt{Hyb-sync} (right) for a
    5-point stencil in two dimensions. We start indexing with 1 and rely on a
    function \texttt{index} flatting arrays. \texttt{m} is the cell data array.
    \label{algorithm:hybrids}
  }
  \begin{minipage}{0.49\textwidth}
  \begin{algorithmic}[1]
    \For{$c \in \{0,1\}$}
      \For{$1\leq y \leq N$}
       \For{$((y+c)\ mod\ 2) + 1 \leq x \leq N; x \gets x+2$}
        \State \#pragma omp task firstprivate(x,y)   
                                 depend(out:m[index(x,y)])   
                                 depend(in:m[index(x+1,y)]) 
                                 depend(in:m[index(x-1,y)]) 
                                 depend(in:m[index(x,y+1)]) 
                                 depend(in:m[index(x,y-1)])
        \State 5-point-stencil(x, y)
       \EndFor
      \EndFor
    \EndFor
    \State \#pragma omp taskwait
  \end{algorithmic}
  \end{minipage}
  \begin{minipage}{0.49\textwidth}
  \begin{algorithmic}[1]
    \For{$c \in \{0,1\}$}
     \State \#pragma omp parallel for
     \For{$1 \leq y \leq N$}
      \For{$((y+c)\ mod\ 2)+1 \leq x \leq N, x \gets x+2$}
       \State \#pragma omp task firstprivate(x,y)
       \State 5-point-stencil(x, y)
      \EndFor
     \EndFor
     \State \#pragma omp taskwait
    \EndFor
  \end{algorithmic}
  \end{minipage}
\end{algorithm}

Graph-based parallelisation yields---also through work stealing---better
balanced work distributions than nested dissection and colouring that relies solely on a \texttt{parallel for}
applied to the outermost sweep loop.
Yet, we find it perform worse if the
computational load per cell is not significant.
Furthermore, its excellent concurrency is consumed by a thread
synchronisation/scheduling overhead that grows with the problem size.
It often yields familiar work assignment patterns.

As multithreading overhead is not negligible for low arithmetic intensity and
very small problems, it might here make sense to refrain from any
parallelisation.
Otherwise, we know that colouring is a
good choice though it lacks the flexibility of the tasking, if the arithmetic
intensity is small and we run on a multicore system.
This motivates us to propose two realisation variants
(Alg.~\ref{algorithm:hybrids}):

In a first approach, we stick to dependency graphs
but issue all tasks according to the colouring pattern (Alg.~\ref{algorithm:hybrids}).
We label this approach as \texttt{Hyb-depend} as it is a hybrid.
While eventually all tasks with their dependencies are passed to the scheduler,
handling the first $N^d/c$ does not require any waits for other active tasks.
Once the first batch of tasks is processed, the likelyhood is high that also
follow-up tasks in the task queue are not blocked by other running tasks.
We can read our traversal as an dependency-based approach that provides
hints to the scheduler what good task orderings might look like.
This approach pays off for very small problem sizes and very sparse
dependencies only.
Notably for dynamically changing, block-structured grids, this is an important
use case.

For our second approach, we found it advantegous to
stick to the colouring but to replace the loops within one colour by tasks.
These tasks are, by construction, dependency free.
BSP's implicit synchronisation is replaced by an explicit
global synchronisation wrapping up each colour.
We can read read our traversal as a colouring approach into which we inject the
work stealing of task-based parallelisation.
We label it as \texttt{Hyb-sync} as it is a hybrid.

%% file: 09_conclusion.tex
\section{Conclusion}
\label{section:conclusion}

Our experiments reveal that OpenMP's both tasking and dependency graph
parallelism are mature though they are, as stand-alone approach, not the
method of choice for the present studies.
Notably, classic colouring typically yields faster codes.
Nested dissection induces fewer global synchronisation points than colouring and
also optimises automatically w.r.t.~caches as it introduces loop tiling. 
The latter property does not materialise here as our algorithm does not reuse
data and, thus, nested dissection is not competitive.
These observations plus an analysis of the arising patterns motivate us to
suggest two hybrid realisations of simple stencil loops that outperform the
naive realisations.
They anticipate that the graph-based approach tends to yield an execution
pattern similar to colouring, i.e.~it seems that colouring's ordering is a
brilliant first guess for a proper task distribution, while the graph modelling injects the amount of flexibility
into the computation that allows OpenMP to cope with heterogeneous load.

Pragma-based parallelism is particular attractive to developers if the baseline
code is not to be altered.
Colouring requires users to decompose the two or three nested loops into three
or four, respectively.
They then are subject to pragmas.
Our proposed hybrid approaches adhere colouring's simplicity.
For the future, it however would be best if users could rely solely
OpenMP's dependency graph scheduler, i.e.~use \texttt{Hyb-depend}, but instruct
the runtime system to start with a colouring-like task distribution
pattern as first guest.
This would simplify codes further.
To harvest asynchronisity, it would be beneficial to be able to specify after
how many tasks in a row natural candidates for synchronisation points arise.